\documentclass[twocolumn,prl,aps,showpacs,superscriptaddress,floatfix]{revtex4}

\usepackage{graphicx}
\usepackage{dcolumn}
\usepackage{bm}
\usepackage{hyperref}
\usepackage{amsmath}

\begin{document}


\title{Microwave Resonance of 2D Wigner Crystal around integer Landau fillings}

\author{Yong P. Chen}
\affiliation{National High Magnetic Field Laboratory, 1800 E.~Paul Dirac Drive, 
Tallahassee, FL 32306}
\affiliation{Department of Electrical Engineering, Princeton University,
Princeton, NJ 08544}
\author{R. M. Lewis}
\affiliation{National High Magnetic Field Laboratory, 
1800 E.~Paul Dirac Drive, Tallahassee, FL 32306}
\affiliation{Department of Electrical Engineering, Princeton University,
Princeton, NJ 08544}
\author{L.\ W.\
Engel}
\affiliation{National High Magnetic Field Laboratory, 1800 E.~Paul Dirac Drive, 
Tallahassee, FL 32306}
\author{D.\ C.\ Tsui}
\affiliation{Department of Electrical Engineering, Princeton University,
Princeton, NJ 08544}
\author{P.\ D.\ Ye}
\altaffiliation[Current address: ]{Agere Systems, 555 Union Blvd., Allentown PA 18109}
\affiliation{National High Magnetic Field Laboratory, 1800 E.~Paul Dirac Drive, 
Tallahassee, FL 32306}
\affiliation{Department of Electrical Engineering, Princeton University,
Princeton, NJ 08544}

\author{L.\ N.\ Pfeiffer}
\affiliation{Bell Laboratories, Lucent Technology, Murray Hill, NJ 07974}
\author{K.\ W.\ West}
\affiliation{Bell Laboratories, Lucent Technology, Murray Hill, NJ 07974}

\date{\today}

\begin{abstract}
We have observed a resonance in the real part of the 
finite frequency diagonal conductivity 
using microwave absorption measurements in high quality 2D electron systems 
near {\em integer fillings}. The resonance exists in some neighborhood of 
filling factor around corresponding integers and is qualitatively similar to 
previously observed resonance of weakly pinned Wigner crystal 
in high $B$ and very small filling factor regime. Data measured around both 
$\nu=1$ and $\nu=2$ are presented. 
We interpret the resonance as the signature of Wigner crystal state around 
integer Landau levels.

\end{abstract}

\maketitle

Two dimensional electron systems(2DES)\cite{2d} 
subjected to perpendicular magnetic 
field $B$
have been observed to display a remarkably rich array of phases in different 
regimes of filling factor $\nu=nh/eB$, where $n$ is the 2D electron density.
These phases include the renowned 
integer and fractional quantum Hall effects (QHE)
\cite{qhereview}, discovered more than two decades ago.
The integer quantum Hall effect (IQHE), 
with $\nu$ taking integer values, has been explained by a disorder
induced \textit{one-particle} localization mechanism; whereas 
the fractional quantum Hall effect (FQHE), with $\nu$ being certain fractions, 
is strictly a many-particle phenomenon. 
At small $\nu$, following the termination of FQHE series in 
the lowest Landau level (LLL), 
the ground state for a sufficiently clean system is believed 
to be a Wigner crystal(WC)\cite{lozoviketc,willet}
and disorder would pin the 
crystal, rendering it insulating\cite{wcreview}. 
One of the experimental supports for such Wigner crystal phase in 
LLL is the recent observations of sharp 
resonance in the real part of frequency($f$) dependent 
diagonal conductivity Re[$\sigma_{xx}(f)$] measured by   
microwave absorption \cite{wcres,peide}.
	
	In this letter we report the observation of similar 
resonances around {\em integer} Landau fillings, which  
we interpret as also coming from a Wigner crystal 
state, formed around integer fillings by electrons/holes in the
top Landau level. This also indicates that pinning of 
a \textit{many-particle} ground state, such as Wigner crystal, 
can be relevant even for IQHE. 

\begin{figure}[tb!]
\includegraphics[height=7cm,width=8cm]{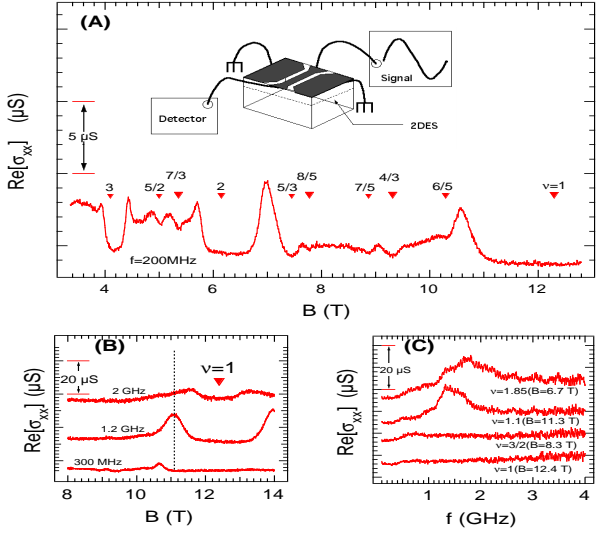}
\caption{\label{fig:fig1} {\bf (A)} The $B$-dependent conductivity
at 200 MHz and $\sim$80 mK with a slightly elevated microwave power. 
Several filling factors are marked.  
Inset shows the schematic measurement circuit. Dark 
regions represent the metallic films deposited on the sample to 
make the CPW. {\bf (B)} $B$-dependent 
conductivity around $\nu=1$ measured at three different frequencies as labeled.
Traces appropriately offset vertically for clarity.
The temperature during $B$-sweep is about 80 mK. Dotted line is a 
guide to the eye for the resonating behavior. 
 {\bf (C)} A few $f$-dependent 
conductivity spectra (offset for clarity) measured at $\sim$50 mK 
and at various $B$ fields labeled underneath each trace. 
Data in both panel B \& C are measured with a low microwave power.}   
\end{figure}

	We have performed our study using  
high quality 2DES in a GaAs/AlGaAs quantum well(QW) structure
grown by molecular beam epitaxy.
The QW is 300 ${\rm\AA}$ wide and located 2000 $\mathrm{\AA}$ 
beneath the surface. The 2DES has as-cooled density 
$n=3.0 \times 10^{11}\ {\rm cm^{-2}}$ and 0.3~K mobility   
about $2.4 \times 10^{7} \ {\rm cm^{2}\ V^{-1}s^{-1}}$. 
A metal film coplanar waveguide (CPW)\cite{engel} of straight line shape 
was patterned onto the
sample surface by photolithography. A schematic of the sample and measuring 
circuit is shown in the inset of Fig.~\ref{fig:fig1}(A).
A network analyzer generates a microwave signal
propagating along the CPW, which couples capacitively to the 2DES, then 
measures the relative power transmission $P$. 
The real part of the diagonal conductivity,   
Re[$\sigma_{xx}$], simply referred to as 
``conductivity'' hereafter, can be related to $P$ as 
Re[$\sigma_{xx}$] $=-\frac{w}{2 l{\rm Z}_{0}}\ln |P/P_0|$
, where $l=2 \mathrm{mm}$ is the length of the CPW, 
$w=20 \mu\mathrm{m}$ its slot width, 
$Z_0=50 \Omega$ is the line impedance in the absence of 2DES and $P_0$ is the  
power transmission in the limit of vanishing 2DES conductivity\cite{engel}. 
Normalizing $P$ by $P_0$ gets rid of the non-flat 
$f$-dependence of microwave attenuation not associated with 
the 2DES, mainly that due to the coaxial cable. 
While the total depletion of 2DES is not possible for our sample, 
$P_0(f)$ is estimated from the power transmission measured under certain 
reference condition, such as at IQHE minimum, at half filling (for example 
$\nu=3/2$), at high power, or at high temperature. Specific examples of 
references are given below. We have used different references to check
our experimental findings and found negligible difference in the results.

	In Fig.~\ref{fig:fig1}(A) we display an example of $B$-dependent
conductivity measured at a fixed frequency (200 MHz) and $\sim$80 mK. 
We can readily resolve such FQHE 
states at 6/5, 4/3, 7/5, 8/5, 5/3, 7/3 and 5/2, attesting to the high 
quality of the sample.  Panel B shows the $B$-dependent conductivity 
measured at three different frequencies, 
in field range of 8 to 14 T, through $\nu=1$.
We see the peak-like ``wing'' (for example the one near 11T in the middle 
trace) 
on the side of $\nu=1$ has a small amplitude 
in the 300 MHz trace (bottom), but is greatly enhanced in the 1.2 GHz trace 
(middle), and reduces to small amplitude again in the 2 GHz trace (top), 
thus displaying a resonating behavior. Such behavior is most clearly
seen through the $f$-dependent conductivity spectrum Re[$\sigma_{xx}(f)$].
Panel C shows
four spectra measured at $\nu$=1, 3/2, 1.1 and 1.85 (from bottom to top) 
respectively, all 
acquired at about 50 mK and in the low microwave power limit. 
The reference spectrum is taken at a much higher power at $\nu=1$.   
Both spectra at $\nu=1$ and $\nu=3/2$ 
are flat within experimental 
tolerance and can actually be used as alternative references($P_0$), 
giving at most a constant offset in the conductivity obtained but 
otherwise having little influence in the results. 
In contrast, the spectrum at $\nu=1.1$($B$=11.3T) shows 
a strong resonance (near 1.3 GHz) of
height more than 10 $\mu S$ and quality factor $Q$ (peak frequency divided by 
FWHM(full width at half maximum)) almost 3. 
Similarly, the spectrum at $\nu=1.85$($B$=6.7T) 
also displays a strong resonance, near 1.7 GHz. 

\begin{figure}[tb!]
\includegraphics[width=8.5cm]{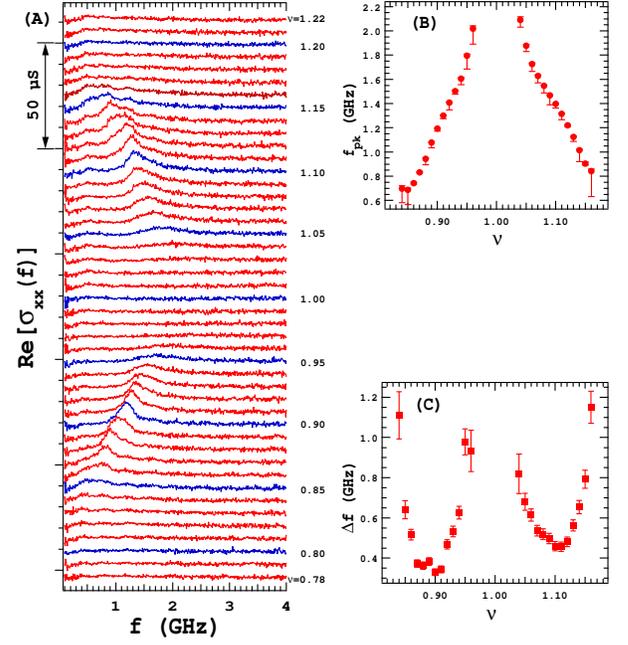}
\caption{\label{fig:fig2}{\bf (A)} 
Frequency dependent conductivity
 spectra around $\nu=1$: from $\nu=0.78$ (bottom trace) to $\nu=1.22$ 
(top trace). Adjacent traces differ 0.01 in $\nu$ and are 
offset 6 $\mu$S from each other for clarity. Filling factors for selected
traces are labeled at right.  Measurements are performed at $\sim50$ mK. 
{\bf (B)} $f_\mathrm{pk}$ versus filling. {\bf (C)} $\Delta f$
versus filling factor.}
\end{figure}

\begin{figure}[tb!]
\includegraphics[width=8.5cm]{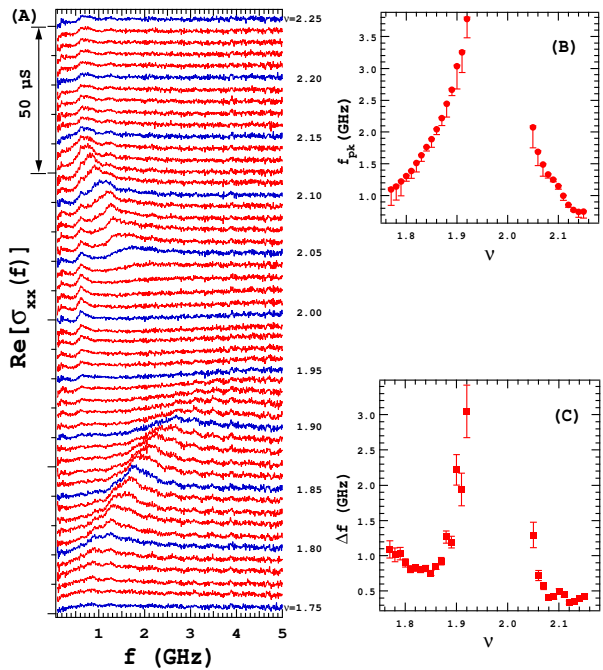}
\caption{\label{fig:fig3}{\bf (A)} 
 Frequency dependent conductivity
spectra around $\nu=2$: from $\nu=1.75$ (bottom trace) to $\nu=2.25$ 
(top trace). Adjacent traces differ 0.01 in $\nu$ and are 
offset 4 $\mu$S from each other for clarity. Filling factors for selected
traces are labeled at right.  Measurements are performed at $\sim$50 mK.
 (The small spike near 600 MHz in some traces, not moving with $B$,
is likely due to an experimental artifact).
 {\bf (B)} $f_{\rm{pk}}$ versus $\nu$. {\bf (C)} $\Delta f$ versus $\nu$.}
\end{figure}

\begin{figure}[tb!]
\includegraphics[width=8.5cm]{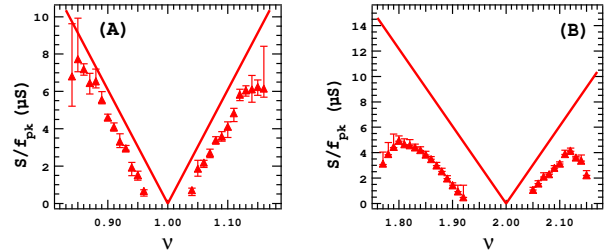}
\caption{\label{fig:fig4}{\bf (A):} 
Oscillator strength divided by peak frequency($S/f_\mathrm{pk}$)
 as a function of
filling factor, for resonance around $\nu=1$. Thick solid line is the 
Fukuyama-Lee result for WC of effective filling factor $\nu^*=\nu-1$.
{\bf (B):}$S/f_\mathrm{pk}$ as a function of
filling factor, for resonance around $\nu=2$.  Thick solid line is the 
Fukuyama-Lee result for WC of effective filling factor 
$\nu^*=\nu-2$. See text for details.}
\end{figure}

	This resonance in $f$-dependent conductivity
happens for $\nu$ near 
integers and has been observed around $\nu=1$, 2, 3 and 4.  
In this paper we focus mostly on the resonances around
$\nu=1$ and $\nu=2$. The resonances around $\nu=3$ and 4 behave similarly
to those around $\nu=1$ and 2, but are much weaker and will be treated in 
detail in a future publication.  All data shown in 
Figures.~\ref{fig:fig2} to \ref{fig:fig4} below 
are measured on an adjacent sample 
from the same wafer, at $\sim$50 mK (except otherwise noted), 
and in the low power limit.
 
	Fig.~\ref{fig:fig2}(A) shows Re[$\sigma_{xx}(f)$] spectra measured at 
45 filling factors ranging from 0.78 to 1.22, in equal
increments of 0.01.
When $\nu$ is sufficiently far from 1 ($\sim$0.8 and 
1.2) the spectrum is flat with no resonance. A resonance starts
to develop when $\nu$ is around 0.84-0.85 (for $\nu$ below 1) and 
1.15-1.16 (for $\nu$ above 1) at frequencies below 1 GHz. 
The resonance sharpens with increasing peak frequency 
as $\nu$ approaches 1 (from both sides) 
till becoming sharpest around $\nu=0.9$ (resonating around 1.2 GHz) and 
$\nu=1.1$ (resonating around 1.4 GHz). As $\nu$ 
further approaches 1 the resonance weakens but its peak frequency continues 
to increase; the last visible resonance is around $\nu=$0.95-0.96 
and 1.04-1.05 with frequency
reaching nearly 2 GHz. In the immediate vicinity of 
$\nu=1$ ($0.96<\nu<1.04$) the spectra are 
again flat. In panel B and C 
of Fig.~\ref{fig:fig2} we plot the peak frequency, 
$f_{\rm{pk}}$, and the full width at half maximum, $\Delta f$,  
of the resonance as functions of 
$\nu$. Here $f_{\rm{pk}}$ and $\Delta f$ are extracted by fitting the 
resonance to a Lorentzian: $A_0+A_1/(A_2+(f-f_\mathrm{pk})^2)$, with
$\Delta f=2\sqrt{A_2} $. While 
$f_\mathrm{pk}$ monotonically increases as $\nu$ moves closer to 1, $\Delta f$
reaches minima when $\nu$ is about 0.1 away from 1, where the 
resonance has quality factor $Q=f_\mathrm{pk}/\Delta f$ of more than 
3.

	In Fig.~\ref{fig:fig3}(A) we display Re[$\sigma_{xx}(f)$] spectra 
measured at 51 filling factors between 1.75 to 2.25, 
again in 0.01 increment of $\nu$. 
The strongest resonance on each side of $\nu=2$ occur at $\nu\sim1.85$ (with 
peak frequency about 1.8 GHz) and $\nu\sim2.12$ (resonating at below 
1 GHz); the peak frequency of the resonance always increases as $\nu\to2$. 
The qualitative features of the resonance are similar
to those around $\nu=1$; 
however we notice an evident asymmetry
between the two sides of $\nu=2$, possibly related to the different wave
functions in different orbital Landau levels (whereas both sides of
$\nu=1$ belong to the same orbital Landau level).  The $f_\mathrm{pk}$  and
$\Delta f$ of the resonance are extracted as in Fig.~\ref{fig:fig1} and 
plotted in panel B and C respectively. 

	The most natural interpretation of our data is that the 
resonance we observe is due to a Wigner crystal phase 
formed around integer Landau fillings.
For clean enough 2DES, Wigner crystal has been theoretically  
assumed to be the ground state of the system for filling factor 
$\nu=K+\nu^*$ with sufficiently small $|\nu^*|$, 
where $K$ is some positive integer\cite{wchllth}.  
Such considerations are often based on the simple physical 
picture that electrons (or holes, for negative $\nu^*$) in filled 
Landau levels can be assumed to be ``inert'' and the remaining 
electrons/holes of ``effective filling factor'' $\nu^*$ and 
density $n^*=(n/\nu)\nu^*
=n\nu^*/(K+\nu^*)$ should Wigner-crystallize when the size of their localized 
wave function (on the order of the magnetic length $l_{B}=\sqrt{\hbar/eB}$) 
becomes small compared to their average 
spatial separation. Due to interaction with weak disorder such a crystalline 
phase is pinned, rendering the top Landau level insulating, 
 and supports a pinning mode\cite{fulee} that gives rise to the observed 
resonance.   

	Our resonance is qualitatively 
similar to the resonance previously observed at small filling factors 
in the LLL Wigner crystal regime 
of both electrons and holes\cite{wcres,peide,ccli}, 
as well as recently discovered resonance
from the ``bubble'' crystal 
phase in high ($\nu>4$) Landau levels\cite{bubble}, all
thought to be caused by the pinning mode of  
crystalline domains in the 2DES oscillating in impurity pinning potential. 
The many-particle nature of such a pinning mode is  
reflected in several features of our observed resonance. For example,
the resonance at 14 T, $\nu= 0.89$ is observed up to nearly 200 mK, 
much higher than the temperature scale corresponding to its resonating 
frequency($\sim$1 GHz$\sim$50 mK). This rules 
out the picture of individual particles 
trapped by disorder giving the resonance.
Furthermore, the resonance (at lowest temperature) can have quality factor 
$Q$ more than 3. The collective motion of a large region of 
particles can average disorder and allow such high $Q$
\cite{ff}.

	Additional insights about our observed resonance in support of 
the pinned Wigner crystal picture can be gained 
by integrating the spectrum to extract the 
oscillator strength $S$\cite{oscdef}.
Previous theory by Fukuyama and Lee gives \cite{fulee,foglersay}
$S/f_{\rm{pk}}=|n^*e\pi/2B|=|(e^2\pi/2h)\nu^*|$ for pinned 
2D WC with density $n^*$
and effective filling factor $\nu^*$ under perpendicular $B$. 
Fig.~\ref{fig:fig4} (A) and (B) 
display $S/f_{\rm{pk}}$ calculated for the resonance in Fig.~\ref{fig:fig2}
(around $\nu=1$) and Fig.~\ref{fig:fig3} (around $\nu=2$) respectively. 
We see indeed the data follow a straight line over most of its range. We 
remark here that the magnitudes of the resonance and $S$ are found 
to have some significant variations (up to factor of 2) 
for different cool-downs and samples (from the same wafer), but such 
linear-like behavior in $S/f_\mathrm{pk}$ versus $\nu^*$ is always observed.

	An important feature of our resonance, as already seen in 
Fig.~\ref{fig:fig2}(B) and Fig.~\ref{fig:fig3}(B), is  
that the peak frequency of the resonance always monotonically 
increases with decreasing effective density $n^*$. This is 
a key character of the ``weak-pinning'' picture\cite{ccli,bvarnote}. 

  From the data presented in this paper, we notice that 
the observed resonance is mostly visible in a $\nu^*$ range of 
$\nu^*_l<|\nu^*|<\nu^*_u$ 
(this range depends on the specific integer filling 
and even the sign of $\nu^*$).  The existence of an ``upper limit'',
$\nu^*_u$ , analogous to the case in LLL Wigner crystal, is probably 
because WC is not the ground state of our 2DES at large enough $\nu^*$.
For example, away from $\nu=1$, as $\nu\to4/5$ or $\nu\to 6/5$,
the system would enter the FQHE state, which is an incompressible liquid.
This kind of ``quantum melting'' of the WC state 
would account for the observed weakening of the 
resonance and 
the drop of $S/f_\mathrm{pk}$ at large $|\nu^*|$ seen in Fig.~\ref{fig:fig4}.
The existence of a ``lower limit'', 
$\nu^*_l$, also corresponding to a ``lower limit'' for density 
$n^*_l=(n/\nu)\nu^*_l$, possibly indicates the ``carriers'' (electrons/holes)
are individually localized by disorder for 
densities $n^*$ below $n^*_l$.
This lower limit of $n^*$ would also imply that higher density samples 
may allow such resonance to be observable around higher integer fillings. 
Preliminary studies performed on a lower density sample
($7\times10^{10}\mathrm{cm}^{-2}$ with mobility 
$\sim 5 \times 10^{6} \ {\rm cm^{2}\ V^{-1}s^{-1}}$) 
have only found relatively weak 
resonance around $\nu=1$ and none around higher integer 
fillings.

	For more disordered 2DES, theories of 
frequency-driven variable range hoping conduction in IQHE\cite{hopth}
predict a linear dependence of  Re[$\sigma_{xx}$] on frequency, which has been 
confirmed in recent experiments\cite{hopexp} on samples of mobilities 
up to $5\times 10^{5} \ {\rm cm^{2}\ V^{-1}s^{-1}}$. No resonance were
seen in these experiments. 
	
	Around $\nu$ = 1 a ``Skyrme'' crystal has been 
proposed\cite{skycth} and there are some 
recent experiments hinting its existence\cite{skycex}. 
We note that the ``Skyrme'' crystal cannot explain our observed
resonance around $\nu=2$, where  
Skyrmions do not exist. 
Moreover the resonance we observe 
shows major similarities around $\nu=1$ and 2
in contrast to the experiments in \cite{skycex}, both of which showed  
a response in their measured quantity that has orders-of-magnitude 
difference between near $\nu=1$ and 
near $\nu=2$.   

In summary, we have observed a microwave resonance around integer Landau 
fillings ($\nu=K+\nu^*$) in high quality 2DES.
At either side of each integer filling $K$ 
with such observable resonance, the peak frequency monotonically increases   
with decreasing $|\nu^*|$, whereas the resonances are strongest at certain 
fillings away from $K$.  We interpret the resonance as 
caused by the pinning mode of a Wigner crystal phase of density 
$n^*=(n/\nu)\nu^*$ formed by 
electrons/holes in the top Landau level, around the corresponding 
integer fillings.


\begin{thebibliography}{}

\bibitem{2d} T.\ Ando, A.\ B.\ Fowler and F.\ Stern, Rev.\ Mod.\ Phys.\ {\bf 54}, 437 (1982).



\bibitem{qhereview}For reviews on QHE, see {\em The Quantum Hall Effect}, 
edited by R.~E.~Prange and S.~M.~Girvin (Springer-Verlag, New York, 1990); 
{\em Perspectives in Quantum Hall Effects}, edited by S.~Das Sarma and A.~Pinczuk (Wiley and Sons, New York 1997).


\bibitem{lozoviketc} Y.~E.~Lozovik and V.~I.~Yudson, JETP Lett., {\bf 22}, 11 (1975); Pui K.~Lam and S.~M.~Girvin, Phys.~Rev.~B., {\bf 30}, 473(1984); 
D.~Levesque, J.~J.~Weis and A.~H.~MacDonald, Phys.~Rev.~B., {\bf 30}, 
1056 (1984); X.~Zhu and S.~G.~Louie, Phys.~Rev.~B., {\bf 52}, 5863 (1995)

\bibitem{willet} R.~L.~Willett \textit{et al.}, Phys.~Rev.~B., {\bf 38}, 7881 
(1988)
 

\bibitem{wcreview}For reviews on WC, see M.\ Shayegan in {\em Perspectives
in Quantum Hall Effects}, edited by S.~Das Sarma and A.~Pinczuk, (Wiley and
Sons, New York, 1997), Chapter 9 and H.~Fertig, Chapter 5. 

\bibitem{wcres}C.~C.~Li \textit{et al}, Phys.~Rev.~Lett. {\bf 79}, 1353 
(1997);L.\ W.\ Engel \textit{et al.}, Solid State Comm. {\bf 104},
167 (1997);L.~W.~Engel \textit{et al.}, Physica E {\bf 1}, 111 (1997)

\bibitem{peide} P.~D.~Ye \textit{et al.}, Phys.~Rev.~Lett. {\bf 89}, 176802 (2002).
 

\bibitem{engel}L.~W.~Engel \textit{et al.}, Phys.~Rev.~Lett. {\bf 71}, 2638 (1993)

\bibitem{wchllth} A.~H.~MacDonald and S.~M.~Girvin, Phys.~Rev.~B {\bf 33}, 4009
(1986);  M.~M.~Fogler, A.~A.~Koulakov and 
B.~I.~Shklovskii, Phys.~Rev.~B {\bf 54}, 1853 (1996); F.\ D.\ M.\ Haldane, E.\ H.\ Rezayi, and Kun Yang,
Phys.\ Rev.\ Lett.\ {\bf 85}, 5396 (2000);
N.\ Shibata and D.\ Yoshioka, Phys.\ Rev.\ Lett.\
{\bf 86} 5755 (2001); N.\ Shibata and D.\ Yoshioka, 
http://xxx.lanl.gov/cond-mat/0210569

\bibitem{fulee}H.\ Fukuyama and P.\ A.\ Lee, Phys.\ Rev.\ B {\bf 17},
535 (1978)

\bibitem{ccli}C.~ C.~ Li \textit{et al.}, Phys.\ Rev.\ B {\bf 61}, 
10905 (2000). 


\bibitem{bubble}R.~M.~Lewis \textit{et al.}, Phys.~Rev.~Lett. {\bf 89}, 136804 (2002)

\bibitem{ff}H.~A.~Fertig, Phys.~Rev.~B. {\bf 59}, 2120 (1999); 
Michael M.~Fogler and David A.~Huse, Phys.~Rev.~B., {\bf 62}, 7553 (2000)

\bibitem{oscdef}Theoretically defined as 
$\int^{\infty}_{0}\mathrm{Re}[\sigma_{xx}(f)]df$. 
For experimental spectrum Re[$\sigma_{xx}(f)$] measured on a 
\textit{finite}  
frequency range $[f_0,f_1]$ and often containing a constant background 
level, we extract $S$ by first computing the indefinite integral $S(f)=
\int^{f}_{f_0}\mathrm{Re}[\sigma_{xx}(f)]df$ for $f_0<f<f_1$, then fitting its 
high frequency part(near $f_e$) to a straight line and subtracting 
the slope from the original Re[$\sigma_{xx}(f)$], then 
recalculating $S(f)$ and taking the difference of its maximum and minimum 
values as $S$.   

\bibitem{foglersay} Michael M.~Fogler, 
private communication, giving
$S/f_\mathrm{pk}$ half the value calculated
in\cite{fulee}.

\bibitem{bvarnote}In our case the $B$ field has 
a slight variation, less than 20\%, around each integer filling.



\bibitem{hopth} A.~L. ~Efros, Sov. ~Phys. ~JETP {\bf 62}, 1087 (1985);
D. ~G. ~Polyakov and B. ~I. ~Shklovskii, Phys. ~Rev. ~B {\bf 48}, 11167 (1993).

\bibitem{hopexp} F.~Hohls, U.~Zeitler and R.~J.~Haug,  Phys. ~Rev. ~Lett. {\bf 86}, 5124 
(2001);R. ~M. ~Lewis and J. ~P. ~Carini, Phys. ~Rev. ~B {\bf 64}, 073310 
(2001).


\bibitem{skycth} L. Brey \textit{et al.}, Phys.~Rev.~Lett. {\bf 75}, 2562(1995);A. G. Green \textit{et al.}, Phys.~Rev.~B {\bf 54}, 16838 (1996); 
R.~Cote \textit{et al.}, Phys.~Rev.~Lett., {\bf 78}, 4825 (1997)

\bibitem{skycex}W.~Desrat {\em et al.}, Phys.~Rev.~Lett. {\bf 88},256807 (2002); 
V.~Bayot \textit{et al.}, Phys.~Rev.~Lett., {\bf 76}, 4584 (1996)

\end{thebibliography}

\end{document}